\documentclass[prd,twocolumn,superscriptaddress]{revtex4}
\usepackage[latin1]{inputenc}
\usepackage{graphicx}
\usepackage{amsmath}
\usepackage{amssymb}
\newcommand{\be}{\begin{equation}}
\newcommand{\ee}{\end{equation}}
\newcommand{\ben}{\begin{eqnarray}}
\newcommand{\een}{\end{eqnarray}}
\newcommand{\bes}{\begin{subequations}}
\newcommand{\ees}{\end{subequations}}

\begin{document}
\title{Generalized Global Defect Solutions}
\author{D. Bazeia} 
\affiliation{Departamento de F\'\i sica, Universidade Federal da Para\'\i ba, 58051-970 Jo\~ao Pessoa, Para\'\i ba, Brazil}
\author{L. Losano}
\affiliation{Departamento de F\'\i sica, Universidade Federal da Para\'\i ba, 58051-970 Jo\~ao Pessoa, Para\'\i ba, Brazil}
\author{R. Menezes} 
\affiliation{Departamento de F\'\i sica, Universidade Federal da Para\'\i ba, 58051-970 Jo\~ao Pessoa, Para\'\i ba, Brazil}
\author{J.C.R.E. Oliveira}
\affiliation{Centro de F\'{\i}sica e Departamento de F\'{\i}sica, Universidade do Porto, 4169-007 Porto, Portugal}

\date{\today}
\begin{abstract}
We investigate the presence of defect structures in generalized models described by real scalar field in $(1,1)$ space-time dimensions.
We work with two distinct generalizations, one in the form of a product of functions of the field and its derivative, and the other as a sum.
We search for static solutions and study the corresponding linear stability on general grounds. We illustrate the results with several examples, where we find stable defect structures of modified profile. In particular, we show how the new defect solutions may give rise to evolutions not present in the standard scenario in higher spatial dimensions.
\end{abstract}
\maketitle
\section{Introduction}

This work deals with defect structures in models described by real scalar field in $(1,1)$ space-time dimensions. These systems support topological and non-topological defect structures which have been studied in diverse scenarios --- see, for instance, Refs.~{\cite{R,VS,MS}}. Usually, in models described by real scalar fields the topological solutions are stable, kinklike defects, and the nontopological solutions are unstable, lumplike defects. An interesting motivation for the study of kinklike defects is that their embedding in four spacetime dimensions gives rise to cosmological domain walls, which should have been formed in primordial phase transitions in the early universe. Domain walls tend to rapidly dominate the energy density of the universe unless they are very light \cite{Zel}, however a number of different domain wall models with interesting cosmological features may be considered. For instance, there are models with non stable domain wall networks where some of the vacua are energetically favoured \cite{n1,n2,n3} and, consequently, the domain wall networks decay during their evolution. Models with non standard domain wall, light enough to satisfy current cosmic microwave background constraints, which may have a possible contribution to dark energy have also been considered \cite{BS} if frozen ({\it frustrated}) networks can be formed \cite{porto}. Another motivation for studyng domain wall networks has recently emerged from the context of brane inflation \cite{bi1}. Therefore, it should be of interest to study the evolution of different domain wall networks in the context of modified dynamics. Other important aspects of studies using scalar field as possible explanations of dark energy concern quintessence \cite{review}, some distinct generalizations of the Chaplygin fluid \cite{cha1,cha2,cha3}, and k-essence \cite{kesse}.

Further extensions come from superstring theories, which have also suggested generalizations of the standard dynamics. An intriguing possibility concerns the tachyon field \cite{S}, in which one modifies the dynamics in a way very similar to the Born-Infeld extension of standard electrodynamics \cite{BI}, done to make the nonlinear contributions to smooth the divergences with appear in the standard case.

The case of a single real scalar field $\phi$ with standard dynamics is governed by the Lagrange density
\be
{\cal L}=X-V(\phi)
\ee
where we are using $X=(1/2)\partial_\mu\phi\partial^\mu\phi$ to represent the kinetic and gradient contributions to the dynamics, and $V(\phi)$ is the potential. Due to results inspired in superstring theories, the scalar field can also evolve under the tachyonic dynamics. In this case, the Lagrange density is modified to \cite{S}
\be
{\cal L}=-V(\phi)\sqrt{1-2X}
\ee

These two distinct possibilities will be used to guide us in the present work, where we introduce and investigate two distinct classes of models. In the first case, we consider extensions of the tachyonic dynamics, that is, we consider models of the type
\be\label{M1}
{\cal L}=V(\phi)F(X)
\ee
where $F(X)$ is in principle an arbitrary function of $X.$ In the second case, we consider extensions of the standard dynamics, that is, we consider models of the form
\be\label{M2}
{\cal L}=F(X)-V(\phi)
\ee 
These two specific classes of models will be investigated below. Similar ideas were already done in Ref.~\cite{babi}, but here we follow another route,
searching for explicit defect solutions and investigating the corresponding classical or linear stability.

The general structure of the present work is organized as follow. In the next Sec.~\ref{scgen}, we present general considerations on the model, and we investigate linear stability of the classical solutions on general grounds. In Sec.~\ref{scmodel} we deal with specific models of both types (\ref{M1}) and (\ref{M2}), searching for static solutions and investigating the corresponding stability. We elaborate on other motivations in Sec.~\ref{3d}, where we investigate features which appear in higher spatial dimensions in clear distinction to the standard scenario.

\section{General considerations}
\label{scgen}

In this section, we turn attention to some characteristics of the general model, which is described by the Lagrange density ${\cal L}(\phi,X).$ Below we deal with the presence of static solutions and the corresponding stability.

\subsection{The model}
\label{scgena}

We consider the case of a single real scalar field. The more general model which preserves Lorentz symmetry is described by the action
\be
{\cal S}=\int \!d^2x \,\,{\cal L}(\phi,X)
\ee
where ${\cal L}(\phi,X)$ represents the Lagrange density, which is to be specified below. We suppose that the Lagrange density does not depend explicitly on the space-time coordinates; thus, the model engenders Poincar\'e symmetry. The equation of motion is given by
\be
\partial_\mu \left({\cal L}_X \partial^\mu \phi\right)={\cal L}_\phi
\ee
where ${\cal L}_X={\partial {\cal L}}/{\partial X}$ and ${\cal L}_\phi={\partial {\cal L}}/{\partial \phi}$. We expand this equation to get
\be\label{eqmotion}
{\cal L}_{X\phi} \partial^\mu\phi \partial_\mu\phi + {\cal L}_{XX} \partial^\mu \phi \partial^\alpha \phi \partial_\mu \partial_\alpha \phi + {\cal L}_{X} \square\phi= {\cal L}_\phi
\ee

We turn attention to the energy-momentum tensor $T_{\mu\nu}.$ It has the form
\be\label{tmn}
T^{\mu\nu}={\cal L}_X \partial^\mu\phi\partial^\nu\phi -\eta^{\mu\nu}{\cal L}
\ee
It is conserved, for field configurations that obey the equation of motion (\ref{eqmotion}). The components are given explicitly by
\bes\label{tmunu}\ben
T^{00}&=&\rho={\cal L}_X\dot\phi^2-{\cal L}\label{00}
\\
T^{01}&=&T^{10}={\cal L}_X\dot\phi\phi^\prime
\\
T^{11}&=&p={\cal L}_X\phi^{\prime2}+{\cal L}
\een\ees
where we are using $\dot\phi=d\phi/dt$ and $\phi^\prime=d\phi/dx.$ 

Since we are dealing with the very general model, we guide ourselves with the null energy condition (NEC), that is, we impose that $T_{\mu\nu}n^\mu n^\nu\geq0,$ where $n^\mu$ is a null vector, obeying $g_{\mu\nu}n^\mu n^\nu=0.$ This condition restricts the model to obey
\be\label{nec}
{\cal L}_X\geq0
\ee 
for $\phi(x,t)$ which solves the equation of motion (\ref{eqmotion}).

We search for defect structures, and we consider static configuration $\phi=\phi(x).$ In this case the equation of motion (\ref{eqmotion})
changes to the simpler form
\be\label{Eq_static_general}
\left(2{\cal L}_{X_sX_s} \,X_s  + {\cal L}_{X_s} \right)\,\phi^{\prime\prime}= 2{\cal L}_{X_s\phi} \,X_s -{\cal L}_\phi
\ee
where we are using the subscript $s$ to remind us to consider static configuration: for instance, $X_s$ stand for $X$ for static field, that is, $X_s=-\phi^{\prime2}/2\leq0.$

We turn attention to the equation of motion (\ref{Eq_static_general}), which can be integrated to give
\be\label{fff234}
{\cal	L}_s - 2{\cal L}_{X_s} X_s =C
\ee
where $C$ is the integration constant. It is interesting to check that this constant $C$ is nothing but the pressure, $T^{11},$ which is constant for static solutions --- see the equations (\ref{tmunu}).

The total energy of the field configuration $\phi(x,t)$ can be obtained as the integral in all space of the energy density, the $T^{00}$ component given in (\ref{00}). If the field configuration describes a static solution we have $\phi=\phi(x)$ and in this case the total energy is given by
\be
\label{ssenergy}
E=-\int^{+\infty}_{-\infty} dx \, {\cal L}(\phi,X_s)
\ee
It identifies the rest mass of the defect structure, and it is important to generalize Derrick's theorem \cite{DH}, to elaborate on the necessary condition for stability of the static solution in the present environment. To do this, we first introduce $\phi^\lambda({x})=\phi(\lambda\,{x}).$ We use $\phi^\lambda$ to define $E_\lambda$ in the form
\be
E_\lambda=- \int^{+\infty}_{-\infty} dx \,{\cal L}\left(\phi^\lambda, -\frac12\left(\frac{d\phi^\lambda}{dx}\right)^2\right)
\ee
We see that $E_{\lambda}|_{\lambda=1}=E$. Thus, we can write
\be
\frac{\partial E_\lambda}{\partial \lambda}=\int^{+\infty}_{-\infty} dx \left(\lambda^{-2}{\cal L}_s- 2\lambda\,{\cal L}_{X_s}X_s\right)
\ee
This expression is to be minimized at the value $\lambda=1$, and this leads to the condition
\be\label{firstorder123}
{\cal L}_s-\,2{\cal L}_{X_s}X_s =0
\ee
We compare this result with (\ref{fff234}) to conclude that only static and pressureless configurations can be stable. 

This equation is a first-order differential equation, since it only depends on the first derivative of the scalar field. This is the pressureless condition, and it is necessary for stability.

\subsection{Linear stability}

The results of the former Sec.~{\ref{scgena}} are very general, and now we complement it with classical or linear stability, which also illustrates how to evaluate quantum effects. We introduce general fluctuations for the scalar field in the form: $\phi(x,t)=\phi(x)+\eta(x,t)$, where $\phi(x)$ represents the static solution. We use these fluctuations in the action to get the quadratic contributions in $\eta$ in the form
\ben\label{s2}
{\cal S}^{(2)}=\frac12\int d^2x \,\{ {\cal L}_X \partial_\mu \eta \partial^\mu \eta +  {\cal L}_{XX} \left(\partial_\mu \phi \partial^\mu \eta \right)^2 \nonumber \\ + \left[{\cal L}_{\phi\phi}-\partial_\mu({\cal L}_{\phi X} \partial^\mu \phi)\right]\eta^2\}
\een
The equation of motion for $\eta$ is then given by
\be\label{eqEstability1}
\partial_\mu \left({\cal L}_X \partial^\mu \eta + {\cal L}_{XX} \partial^\mu \phi \partial_\alpha \phi \partial^\alpha \eta \right)=\left[{\cal L}_{\phi\phi} -\partial_\mu({\cal L}_{\phi X} \partial^\mu \phi) \right]\eta 
\ee 
Since $\phi$ is static solution, we can write
\be
{\cal L}_{X_s} \ddot \eta-\left[\left(2{\cal L}_{X_s X_s}X_s+{\cal L}_{X_s}\right)\eta^\prime\right]^\prime=
\left( {\cal L}_{\phi\phi}+({\cal L}_{\phi X} \phi^\prime)^\prime \right)\eta
\ee
We suppose that  
\be
\eta(t,x)=\eta (x) \cos(\omega t)
\ee
to obtain
\be\label{eqestability2005}
-\left[\left(2{\cal L}_{X_s X_s} X_s+{\cal L}_{X_s}\right)\eta^\prime\right]^\prime\!=\!\left({\cal L}_{\phi\phi}+({\cal L}_{\phi X_s} \phi^\prime)^\prime+\omega^2{\cal L}_{X_s}\!\right)\!\eta
\ee
To ensure hyperbolicity, we impose that \cite{babi}
\be\label{AAAAA}
A^2\equiv\frac{2{\cal L}_{X_sX_s}X_s+{\cal{L}}_{X_s}}{{\cal{L}}_{X_s}} >0
\ee
The above equation (\ref{eqestability2005}) has the form
\be
-[a(x)\eta^\prime ]^\prime = b(x)\eta
\ee
where 
\bes\ben
a(x)&=&2{\cal L}_{X_s X_s} X_s + {\cal L}_{X_s} \\
b(x)&=&{\cal L}_{\phi\phi}+({\cal L}_{\phi X} \phi^\prime)^\prime + \omega^2 {\cal L}_{X_s}
\een\ees

To ease the investigation, we introduce new variables. We make the changes
\ben\label{eqRED}
dx=A\,dz \,\,\,\,\,\,\,{\rm and}\,\,\,\,\,\,\ \eta= \,\frac{u}{\sqrt{{\cal L}_X\,A}}
\een
which allows writing the Schr\"odinger-like equation
\be\label{Aeq1normal2}
- u_{zz}+U(z)\,u=\omega^2 u
\ee
where
\be
U(z)=\frac{\left(A{\cal L}_X\right)^{\frac12}_{zz}}{\left(A{\cal L}_X\right)^{\frac12}}
-\frac{1}{{\cal L}_{X}}\left[{\cal L}_{\phi\phi}\!+\frac1{A}\left({\cal L}_{\phi X}\frac{\phi_z}{A}\right)_z\,\right]
\ee
is the quantum-mechanical potential we have to solve to have the corresponding eigenvalues and eigenstates.

Linear stability requires that the eigenvalues $w^2$ are non-negative, and this crucially depends on the potential $U(z),$ which should be investigated for each one of the specific models that we explore in the next section. An interesting issue concerns the possibility of $w$ being zero, giving as the corresponding eigenstate the zero mode. Although we are considering generalized models, they engender Poincar\'e invariance, and so no defect solution should grant a privilege to localize itself in the real line \cite{jac}: if $\phi(x)$ is defect solution, the infinitesimal translation $\phi(x+\epsilon)=\phi(x)+\epsilon (d\phi/dx)$ should be costless. To quantify the reasoning, we use the quadratic action (\ref{s2}) to obtain the expression
\ben
\frac12\eta\,\{({\cal L}_{\phi X}\phi^{\prime}+2{\cal L}_{XX}X\eta^{\prime}+{\cal L}_X\,\eta^{\prime})^{\prime}+{\cal L}_{\phi\phi}\}\eta
\een
which can be integrated to give the energy contribution. However, we use the equation of motion (\ref{eqestability2005}) with $w\to0$ to see that this quantity vanishes, showing that the zero mode is indeed the derivative of the defect solution, and that it is costless.

\section{Models}
\label{scmodel}

The simplest model is the standard model, which is described by
\be
{\cal L}=X-V(\phi)
\ee
In this case, we have that ${\cal L}_X=1,$ ${\cal L}_{XX}=0,$ ${\cal L}_{X\phi}=0,$ ${\cal L}_\phi=-V_\phi,$ and ${\cal L}_{\phi\phi}=-V_{\phi\phi}.$
We see that the former results reproduce the well-known results of the standard situation. For instance, we use (\ref{Eq_static_general}), (\ref{ssenergy}) and (\ref{eqestability2005}) to get, respectively
\be
\phi^{\prime\prime}=V_\phi
\ee
\be
E=\int dx\left(\frac12\phi^{\prime2}+V(\phi)\right)
\ee
and
\be
-\eta^{\prime\prime}+V_{\phi\phi}\,\eta=w^2\eta
\ee
which we recognize as the equation of motion and energy of the static field, and the Schr\"odinger-like equation for the fluctuation in the standard model.

We notice that for the standard situation, the pressureless condition (\ref{firstorder123}) leads to  
\be\label{Virial}
\frac12 \,\phi^{\prime2} = V(\phi)
\ee
which shows that the gradient and potential energy densities contribute evenly to the total energy of the static configuration.

We use the $\phi^4$ model to represent the standard situation. In this case the potential has the form
\be\label{phi4}
V(\phi)=\frac12(1-\phi^2)^2
\ee
This model is solved by the kinklike defect structures
\be\label{kink}
\phi(x)=\pm\tanh(x)
\ee
The energy density is 
\be
\rho(x)={\rm  sech^4(x)}
\ee
which gives the energy $E=4/3.$ In Fig.~{\ref{blmofig1}} we plot with bold solid line the above potential, defect solution and energy density,
which may be seen as the standard scenario, useful for the extensions to be considered below. In this case, we can use the width of this standard solution as reference for the width of the defect solutions of the extended models of Sec.~\ref{scmodelII}.

The form of the Lagrange density of the standard model suggests that we introduce two distinct classes of models, as we do in the next subsections.

\subsection{Extended models of the first type}

In this case, we concentrate on models described by the Lagrange density (\ref{M1}),
where $V$ and $F$ are function to be specified. In these models, the dynamics is inspired in the tachyon field. The equation of motion (\ref{Eq_static_general}) has the form 
\be
\left(2\,F_s^{\prime\prime}X_s+F_s^\prime\right)\phi^{\prime\prime}=\frac{V_\phi}{V}\left(2\,F_s^\prime X_s-F_s\right)
\ee 
and the first-order equation is
\be\label{eq1oe}
V(\phi)\left(F_s-\,2F^\prime_s X_s\right)=0
\ee
where $F_s=F(X_s)$ and prime means derivative with respect to the argument of the function.

We first consider the case $V=1;$ this leads to a class of models that support no static finite energy field configuration. The next case is for $V(\phi)$ generic; in this case the first-order equation (\ref{eq1oe}) is an algebraic equation, and the solutions describe constant $X_s.$ If these constant values are finite, the field configurations are given by
\be
\phi_i(x)=a_i \,x
\ee
which correspond to $X_s^i=-a^2_i/2,$ such that $X_s=F_s/2F^\prime_s$ are solved by $X_s^i, i=1,2,...$ These topological solutions diverge asymptotically, but the energy may be finite if $V(\phi)$ is properly chosen. In this case, the energy (\ref{ssenergy}) becomes 
\ben\label{ene1}
E_i&=&-F(-a_i^2/2) \int^{+\infty}_{-\infty} \,dx V(a_i\,x)\nonumber\\
&=&-\frac{F(-a_i^2/2)}{|a_i|}\int^{+\infty}_{-\infty}\,d\phi\,V(\phi) 
\een
We notice that all the solutions are pressureless and independent of the specific form of $V(\phi),$ although the energy depends crucially on $V(\phi)$. For this reason, we have to choose $V(\phi)$ properly to make the integral (\ref{ene1}) well-defined. If $V(\phi)$ is non-negative, we have to set 
\be
F(-a_i^2/2)<0
\ee
to make the static solution stable.

Suppose now that $V(\phi)\geq0,$ with zeros at the set $b_i, i=1,2,...$ In this case, another kind of solutions of (\ref{eq1oe}) have the profile
\be\label{sk23323}
\phi_{ij}(x)=\left\{\begin{array}{cc} b_j, &{\rm for} \,\,x>0\\
\frac12(b_i+b_j), &{\rm for} \,\,x=0\\
b_i	, &{\rm for} \,\,x<0 \end{array} \right.
\ee
where $i,j$ label consecutive zeros, with $i<j$ and $b_i<b_j,$ and $F^\prime_s\to0,$ and $X_s\to-\infty,$ such that $F^\prime_sX_s\to F_s/2.$
These configurations have energy density localized at $x=0,$ the center of the kink, and finite energy, given by
\be
E=\lambda \int^{b_j}_{b_i} d\phi\, V(\phi)
\ee
where $\lambda$ is such that, for $X_s\to\infty$ then $F_s\to-\lambda\phi^\prime.$

This type of models have been recently considered for string inspired systems of tachyon condensation \cite{S,mina,brax,hashi}.

We take as an explicit example, the model introduced in Ref. \cite{brax}, in which 
\be\label{func2341io}
F(X)=-(1-2X)^a
\ee
where $a$ is real and positive. The NEC restriction (\ref{nec}) implies that $X<1/2,$ which is satisfied by all static solutions.

The first-order equation (\ref{eq1oe}) gives
\be
V(\phi)(1-2X)^{a-1}\left[1+2(2a-1)X\right]=0
\ee
and now we get the solution $X_s=-{1}/{2(2a-1)}.$ The restriction (\ref{AAAAA}) implies that $a>1/2.$ The static field is given by
\be
\phi(x)=\pm		\frac{x}{\sqrt{2a-1}}
\ee
We notice that the case $a=1$ was investigated in \cite{mina}.
 
The energy has the form
\be
E=(2a)^a (2a-1)^{\frac12 -a} \int^{+\infty}_{-\infty} d\phi \,V(\phi)
\ee

In the above model, the limit $a\to1/2$ is very interesting, since it reproduces the tachyon model considered in \cite{bmr} --- see also Ref.~{\cite{S}}. Here the first-order equation requires that the defect presents singular profile, identified as the singular tachyon kink
\be
\phi(x)=\left\{\begin{array}{cc} \infty, &{\rm for} \,\,x>0\\
0, &{\rm for} \,\,x=0\\
-\infty	, &{\rm for} \,\,x<0 \end{array} \right.
\ee
In the tachyon model, $V(\phi)$ is usually non-negative, and attains its maximum at $\phi=0,$ going to zero asymptotically. Some nice functions are  $V(\phi)={\rm sech}(\phi)/\pi$, $V(\phi)={\rm sech}^2(\phi)/2$ and $V(\phi)=e^{-\phi^2}/\sqrt{\pi},$ which integrate to unit, giving unit energy
to the corresponding defect structure.

\subsubsection{Linear Stability}

Let us turn attention to the classical or linear stability of the model described by (\ref{M1}). The equation for the perturbation (\ref{eqEstability1}) has the form
\be\label{eqEstability122}
\partial_\mu \left[V(F^\prime \partial^\mu\eta\! +\! {F}^{\prime\prime}\partial^\mu\phi\partial_\alpha \phi\partial^\alpha \eta) \right]\!=\!\left[V_{\phi\phi}F\! -\!\partial_\mu(V_\phi F^\prime \partial^\mu \phi)\right]\eta 
\ee 
We consider static solutions to obtain 
\be
\left[V\left(-2F_s^{\prime\prime}X_s\!-\!F_s^\prime\right)\eta^\prime\right]^\prime\! =\!\left[V_{\phi\phi}F_s\!+\! (F_s^\prime V_\phi \phi^\prime)^\prime+\omega^2VF_s^\prime\right]\!\eta
\ee
The use of the equation of motion leads to
\be
\left[V\left(-2F_s^{\prime\prime} X_s\! -\! F_s^\prime \right)\eta^\prime\right]^\prime\! =\!\left[\left(2F_s^{\prime\prime} X_s\! +\! F_s^\prime\right) \phi^{\prime\prime}V_\phi+\omega^2 V F_s^\prime  \right]\! \eta
\ee
For solutions with $X_s$ constant, only the $V$ terms depend on $x.$ Thus, we get
\be
-\frac{A^2}{V}\left( V\eta^\prime\right)^\prime=\omega^2\, \eta
\ee
where $A$ is given by (\ref{AAAAA}) and has the form 
\be
A^2=\frac{2F_s^{\prime\prime} X_s + F_s^\prime}{F_s^\prime}.
\ee
We consider (\ref{eqRED}) and we make the changes
\be
\eta=\frac{1}{\sqrt{V}}\,u  \,\,\,\,\,\,\,\,{\rm and}\,\,\,\,\,\,\,\,x=A \,y
\ee
to get to 
\be
-u_{zz}+ U(z) u = \omega^2 u
\ee
where 
\be
U(z)=\frac1{\sqrt{V}} \left(\frac{1}{\sqrt{V}}\right)_{zz}
\ee
For the function (\ref{func2341io}) we have 
\be
A^2=\frac{2a-1}{a}
\ee	
If we consider the functions $V(\phi)={\rm sech}(\phi)/\pi$, $V(\phi)={\rm sech}^2(\phi)/2$ and $V(\phi)=e^{-x^2}/\sqrt{\pi},$ the quantum mechanical potential are $U(z)=\pi/4[\cosh(z)+{\rm sech}(z)],$ $U(z)=2\cosh(z)^2$ and  $U(z)=\sqrt{\pi}\,e^{-z^2}(z^2+1).$ Their profile are very similar, and  shows that the quantum-mechanical problem supports no negative eigenvalue, ensuring stability of the defect solutions. The zero mode which comes from translational invariance should be $\eta_0={d\phi}/{dx}=c_i,$ but this is constant and non-normalizable. Thus, there is no zero mode, and all the fluctuations are bounded to the defect.

\subsection{Extended models of the second type}
\label{scmodelII}

We consider the second type of models, in which we maintain the potential present in the standard scenario, but change the kinematics.
The general structure of the models is now given by (\ref{M2}), and we suppose that the function $F(X)$ is arbitrary but reproduces the standard structure for $X$ small \cite{babi}. 

In general, the equation of motion is
\be\label{eq2}
\partial_\mu \left(F^\prime \partial^\mu \phi \right)+ V_\phi=0
\ee
We notice that in the standard situation $F(X)=X,$ and $F^\prime=1,$ which makes the above equation the standard equation of motion.
We can rewrite this equation in the form
\be
F^{\prime\prime} \partial^\alpha \phi\, \partial^\mu \phi \,\partial_\alpha \partial_\mu \phi  + F^\prime \square \phi + V_\phi=0
\ee
We now search for static solution, $\phi=\phi(x)$ to get to
\be
(F_s^\prime - F_s^{\prime\prime} \phi^{\prime 2}) \phi^{\prime\prime}=V_\phi
\ee
where $F_s=F(-\phi^{\prime2}/2).$ In this case, the first-order equation has the form
\be\label{derrick111}
 F_s -2\,F_s^\prime X_s = V(\phi) 
\ee

The energy density for the static solution is
\be
\rho=-F\left(X_s\right)+V(\phi)
\ee 
We use (\ref{derrick111}) to obtain
\be\label{eqeqffenrgy}
\rho=F_s^\prime \phi^{\prime2}
\ee
The restriction which comes from the NEC gives $F^\prime\geq0$. Thus, the energy density is non negative, which makes the energy positive definite.

In general, the non-trivial form of (\ref{derrick111}) suggests that it is not always possible to solve this problem analytically. However, we can make further progress supposing that there is a first-order equation of the form
\be\label{firstorder1}
\phi^\prime=W(\phi)
\ee
where $W=W(\phi)$ is a function of the scalar field. This choice imposes that the potential has to have the specific form
\be\label{SSpotential}
V(\phi)=F+F^\prime\,W^2
\ee
where in $F$ and in $F^\prime$ we have to change $X\to-\phi^{\prime2}/2$ and, with the use of the first-order equation (\ref{firstorder1}),
to $-W^2/2,$ which then gives the potential as a function of $\phi.$

We illustrate the general situation with some examples. Firstly we choose the following function
\be\label{example}
F(X)=X+\alpha X^2
\ee
where $\alpha$ in real parameter, which controls the extension of the model. The NEC restriction leads to $1+2\alpha X \geq 0$, and for static solution we get $1-\alpha \phi^{\prime 2} \geq 0.$ The hyperbolicity condition is $1+6\alpha X \geq 0$, and so if this condition is valid, the NEC is also valid. If $\alpha$ is negative, this condition is always valid. If $\alpha$ is positive, we must take
\be
-\frac{1}{\sqrt{3\alpha}} \leq \phi^\prime \leq \frac{1}{\sqrt{3\alpha}}
\ee
We notice that when $\alpha\to0$, we return to the standard case and $\phi^\prime$ is not constrained anymore.

The first order equation can be written as  
\be
\frac12\phi^{\prime 2}-\frac34\alpha\phi^{\prime4}=V(\phi)
\ee
It is solved by
\be\label{goodbad}
\frac12\phi^{\prime2}=\frac{1}{6\alpha}-\frac{\sqrt{1-12\alpha V(\phi)}}{6\alpha}
\ee
The limit $\alpha\to0$ leads to the first-order equation (\ref{Virial}), and this suggests that for the expanded model we define the effective potential in the form
\be\label{veff}
V_{eff}(\phi)=\frac{1}{6\alpha} - \frac{\sqrt{1-12\alpha V(\phi)}}{6\alpha}
\ee 

The form of $F(X)$ is given by Eq.~(\ref{example}), and it leads to the result that if the potential $V(\phi)$ is non-negative, so will be the effective potential in (\ref{veff}). Besides, the zeros of $V(\phi)$ will also be zeros for $V_{eff}(\phi).$ This implies that the topological structure of $V(\phi)$ is preserved in the effective potential $V_{eff}(\phi).$ If $\alpha>0$, the highest value of $V(\phi)$ should be  $1/12\alpha,$ with $V_{eff}(\phi)$ being twice that value, that is, $1/6\alpha.$ 

For kinklike solution connecting two minima of the potential, the center of the kink, which corresponds to the field with highest inclination, is at the maximum of the potential, $V_0,$ in between the two minima. In general, the thickness of the solution depends on the maximum $V_0,$ thus it will certainly be affected by the parameter $\alpha.$

\begin{figure}[ht!]
\centering
\includegraphics[width=6.2cm]{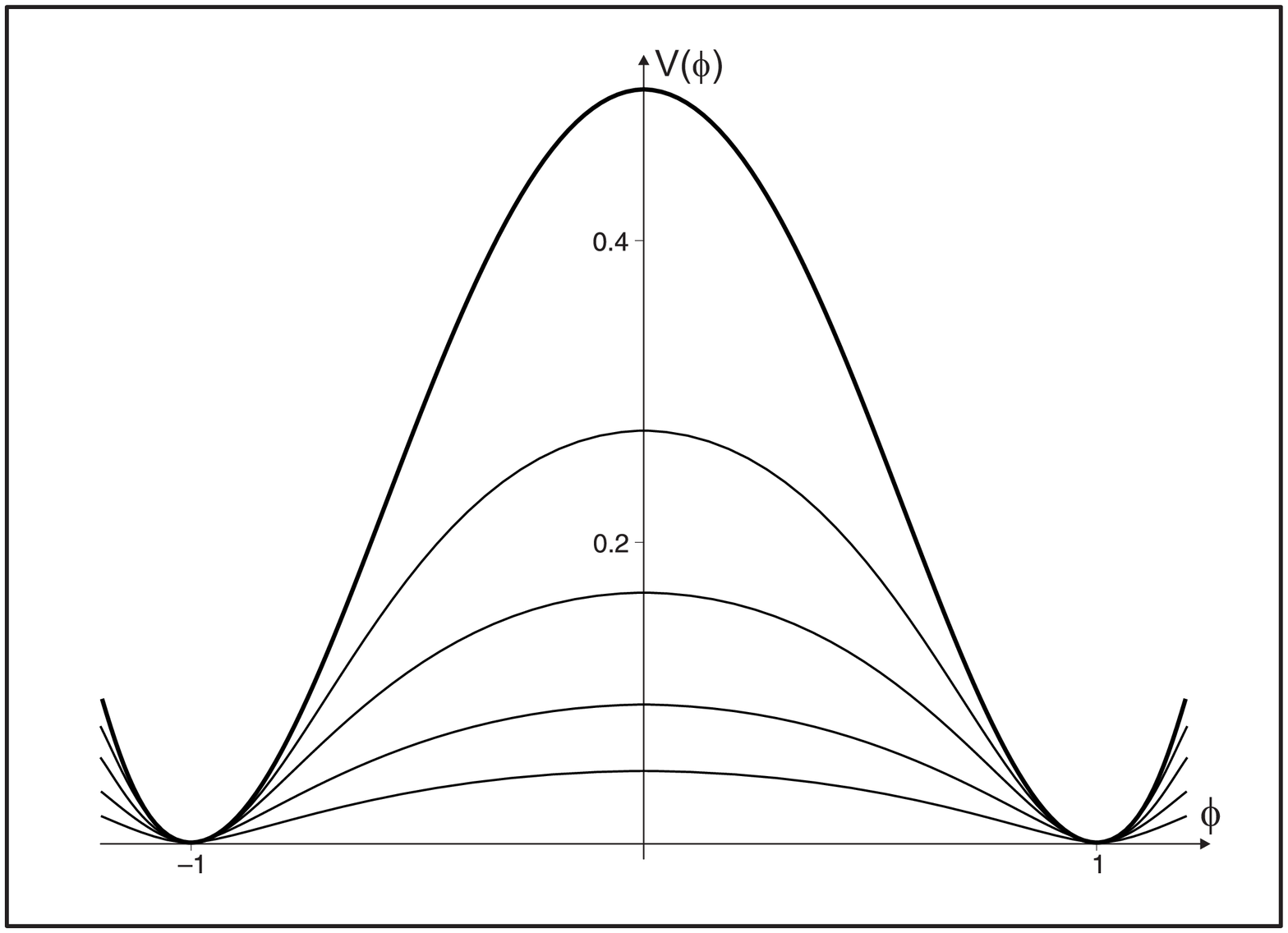}
\includegraphics[width=6.2cm]{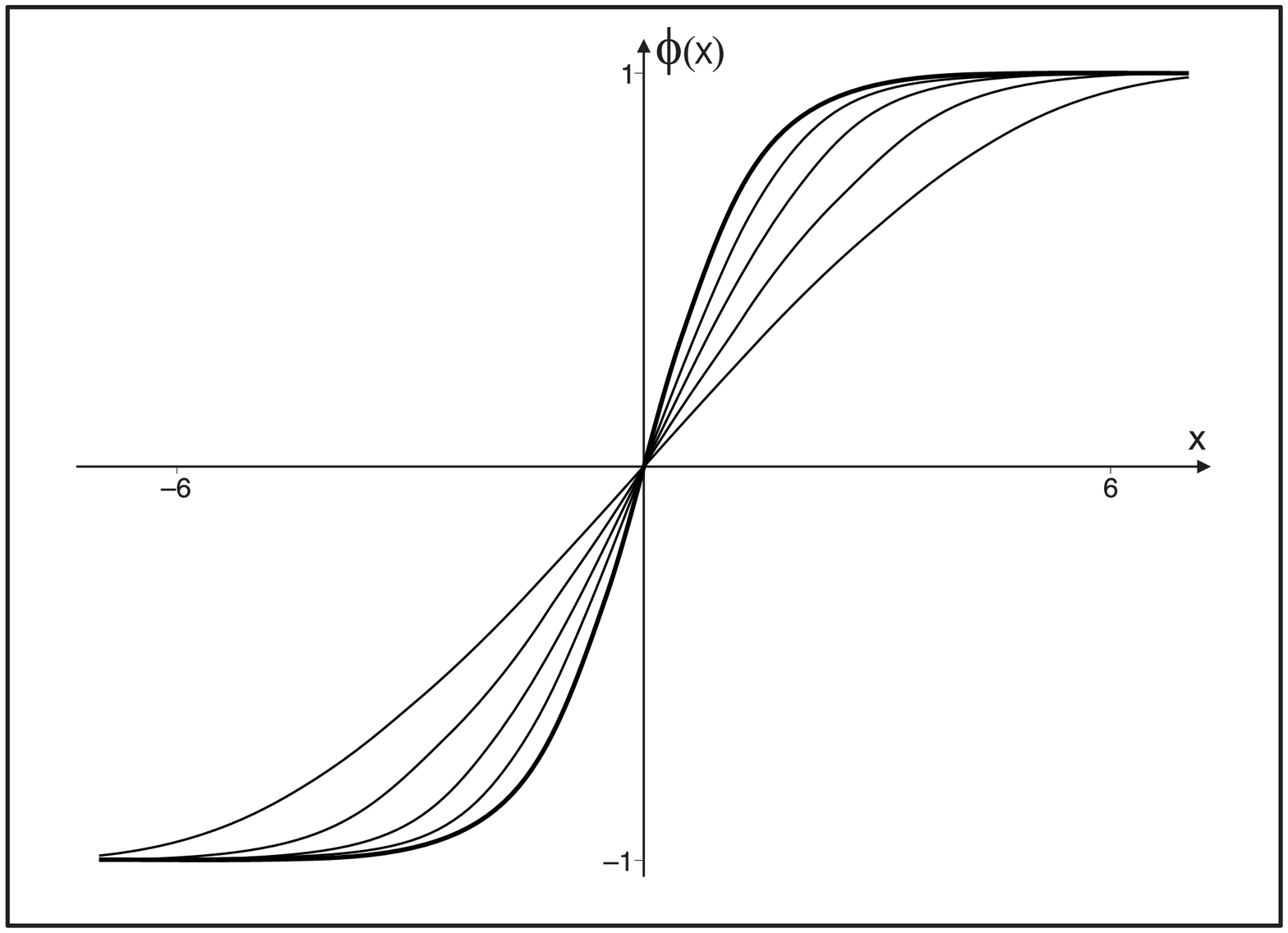}
\includegraphics[width=6.2cm]{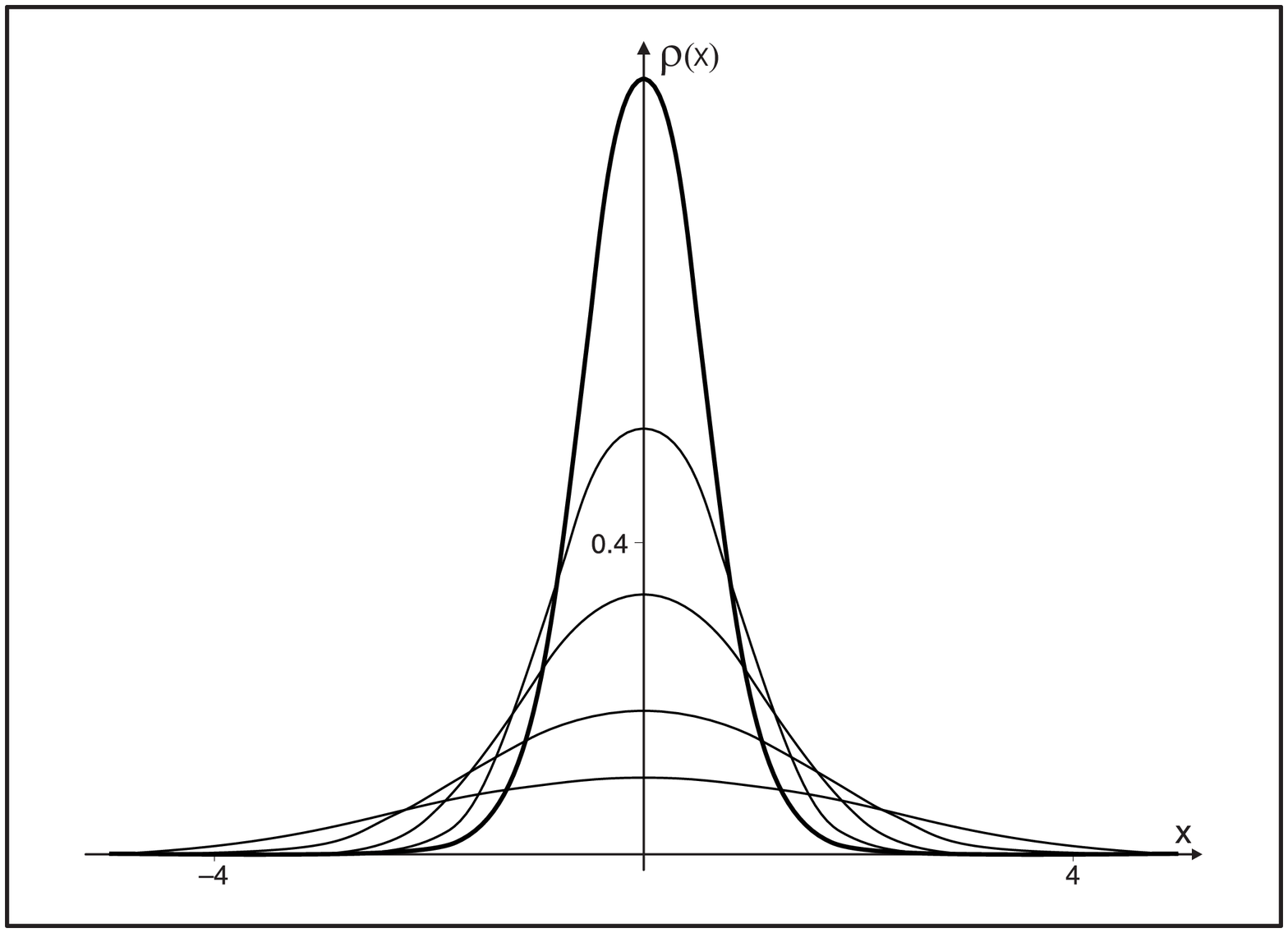}
\caption{Plots of the effective potential (\ref{2de2}) (upper panel), the corresponding defect solutions (middle panel)
and energy densities (lower panel) for $\alpha=0,-1,-4,-16,$ and $-64.$ The bold line is for $\alpha=0;$ the other lines follow
the given sequence.}
\label{blmofig1}
\end{figure}

If the effective potential is chosen as the potential for the standard theory, both standard and extended theories would have the very same defect solutions. In spite of this, both theories are different and would induce distinct behavior for the static solutions: for instance, the energy density
in the standard is given by $\bar\rho=\phi^{\prime2},$ which is different from the value $\rho=\phi^{\prime2}-\alpha \phi^{\prime4}$ of the extended model, which shows that $\rho>\bar\rho$ for $\alpha>0$, and $\rho<\bar\rho$ for $\alpha<0.$ Another distinction concerns stability, and the nonlinear profile of $F(X)$ induces distinct stability behavior on the extended model.

Let us now consider $\alpha$ very small. In this case we can expand the square root to get the effective potential in the form
\be
V_{eff}(\phi)=V(\phi)+3\,\alpha\, V(\phi)^2
\ee
Thus, if we consider $V(\phi)$ polynomial of degree $n$, then the effective potential will also be polynomial, of order $2n.$ For instance, if we choose the $\phi^4$ profile for the potential (\ref{phi4}) we get the effective potential as
\be
V_{eff}(\phi)=\frac12(1-\phi^2)^2+\frac34\alpha(1-\phi^2)^4
\ee
Since $\alpha$ is very small, we can rewrite the above equation (\ref{goodbad}) in the form
\be
\int\frac{d\phi}{\sqrt{2V(\phi)}}-\frac34\alpha\int d\phi\,\sqrt{2V(\phi)}=x
\ee
We see that for $\alpha\to0,$ we get to the standard situation. We now use the $\phi^4$ potential (\ref{phi4}) to get
\be
{\rm arctanh}(\phi)-\frac34\alpha\left(\phi-\frac13\phi^3\right)=\pm x
\ee
and we have being unable to write $\phi=\phi(x).$ However, for $\alpha$ very small we have
\be\label{ssa}
\phi(x)=\pm\tanh\left(x\right) \left(1+ \frac14 \alpha\,{\rm sech}^2(x) \left( 2  + {\rm sech}^2\left(x\right)\right) \right)
\ee
This approximate solution has energy density
\be
\rho(x)={\rm sech}^4(x)\left(1-\alpha\left(2-{\rm sech}^2(x)-\frac32 {\rm sech}^4 (x)\right)\right)
\ee
It can be integrated to give the energy $E=4/3-8{\alpha}/35.$

We now consider $\alpha$ generic, and $V(\phi)$ as the $\phi^4$ model. The effective potential for (\ref{phi4}) becomes
\be\label{2de2}
V_{eff}=\frac{1}{6\alpha}-\frac{\sqrt{1-6\alpha (1-\phi^2)^2}}{6\alpha}
\ee
which we plot in Fig.~{\ref{blmofig1} together with the corresponding defect solutions and energy densities, for the several values
$\alpha=0,-1,-4,-16,$ and $-64.$ We use this figure to notice from the profile of the defect solution that its width and energy changes with $\alpha,$
and this nicely illustrate how the parameter used to extend the model modifies the physical characteristics of the corresponding topological solution.   
We can also choose another model, in which we consider the $\alpha$-dependent $\phi^8$ potential
\be\label{phi8}
V=\frac12(1-\phi^2)^2-\frac34\,\alpha\,(1-\phi^2)^4
\ee
For $\alpha>1/3,$ the potential has seven critical points, given by $\pm1$, $0,$ and $\pm(1\pm(3\alpha)^{-1/2})^{1/2})$. The first three points are minima and the others are maxima. For $\alpha<1/3$, the critical points are $\pm1,$ and $0$. The two first points are minima and the last one is a maximum.

The kinklike solutions are the same of the $\phi^4$ model ---  given in (\ref{phi4}). We use (\ref{eqeqffenrgy}) to find the energy density 
\be
\rho(x)={\rm sech}(x)^4 - \alpha \, {\rm sech}(x)^8
\ee
We consider $\alpha<1$, to make the energy density non-negative. The total energy is $E=4/3-32{\alpha}/35.$ 
In this new model, although the defect solution does not change, the energy density varies with $\alpha$ in an interesting way, as we show in Fig.~{\ref{blmofig2}}. In particular, we notice that as $\alpha$ increases toward unit, the energy density opens an internal gap, in a way similar
to the model investigated in \cite{PRL}, and further considered in \cite{JCAP} as a model which leads to a braneworld scenario with internal structure. This understanding suggests the investigation of the present model in the five-dimensional spacetime with warped $AdS_5$ geometry, but this is out of the scope of the present work and will be further considered elsewhere.

\subsubsection{Linear Stability}

Let us now turn attention to the issue concerning classical or linear stability for the models described by (\ref{M2}). In standard theory, $F(X)=X$ and the kinklike defect structures are stable against small perturbations. To see how the nonlinearity in the kinetic term may influence stability, let us consider the equation (\ref{eqEstability1}) for the fluctuations in the modified model; we get
\be
\partial_\mu\left(F^\prime\,\partial^\mu\eta + F^{\prime\prime}\partial^\mu\phi\partial^\alpha\phi\,\partial_\alpha\eta\right)+V_{\phi\phi}\eta=0
\ee  
where $F^{\prime}$ and $F^{\prime\prime}$ are functions of ${\phi^\prime}.$ For $\phi$ being static solution, we can write
\be\label{eqaaa}
-\left[(F^\prime-F^{\prime\prime} \phi^{\prime2} ) \eta^\prime\right]^\prime+V_{\phi\phi} \eta =\omega^2 F^\prime \eta
\ee
The zero mode solution as in the standard case is the translational mode, $\eta_0=\phi^\prime.$

\begin{figure}[ht!]
\includegraphics[width=7cm]{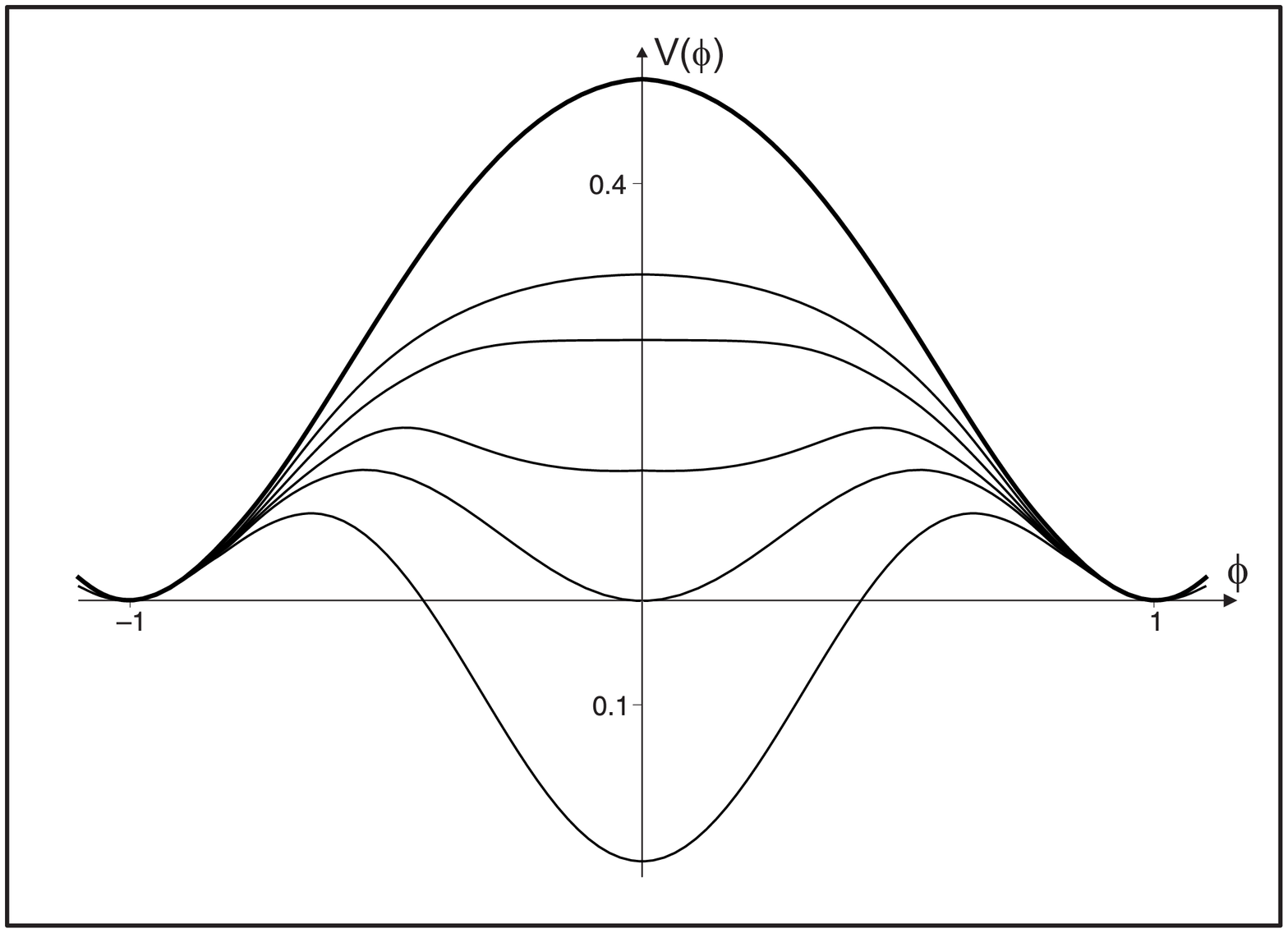}
\includegraphics[width=7cm]{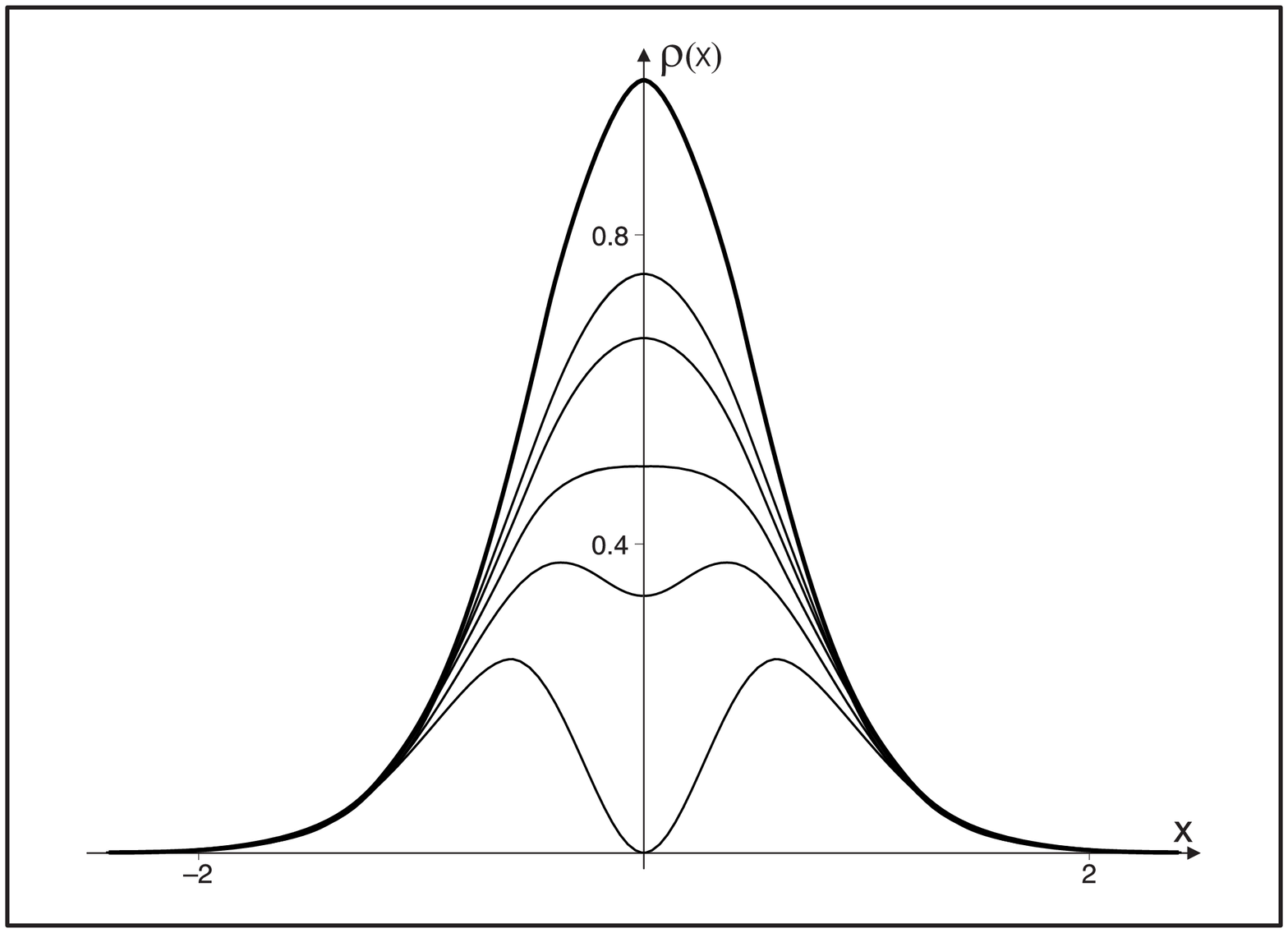}
\caption{Plots of the potential (\ref{phi8}) (upper panel) and the corresponding energy densities for the standard defect solution (\ref{kink}) which solve this model, for the several values $\alpha=0,1/4,1/3,1/2,2/3,$ and $1.$ The bold line is for $\alpha=0;$ the other lines
follow the given sequence.}\label{blmofig2}.
\end{figure}

One possibility appears when the model obeys (\ref{firstorder1}) and (\ref{SSpotential}). In this case, we can rewrite the above equation
(\ref{eqaaa}) in the form
\be\label{eqestab23}
-\left[(F^\prime-F^{\prime\prime}W^2)\eta^\prime\right]^\prime + U\eta=\omega^2F^\prime \eta
\ee
where
\be
U=(W_\phi^2+W W_\phi)(F^\prime-F^{\prime\prime} W^2)+W^2 W^2_\phi\left(F^{\prime\prime\prime}W^2-3F^{\prime\prime}\right)
\ee
Here $F$ should be seen as a function of $\phi,$ like in Eq.~(\ref{firstorder1}). In this case, we can write
\be
S^\dag S \,\eta = \omega^2 F^\prime\, \eta
\ee
with
\be
S=\sqrt{F^\prime-F^{\prime\prime} W^2}\left(-\frac{d}{dx}+W_{\phi}\right)
\ee
We notice that the NEC restriction $(F^\prime>0)$ is important, but we have to impose $F^\prime-F^{\prime\prime}W^2>0$ for the above expression to make sense, and this depends on the specific model under consideration.

We illustrate the situation with the case given by (\ref{example}), and with the potential (\ref{phi4}).
We use (\ref{AAAAA}) to get
\be
A^2=\frac{1+6\alpha X}{1+2\alpha X}
\ee
We suppose that $\alpha$ is very small. In this case, up to first order in $\alpha$ we can write
\be
A=1+2\alpha X=1-\alpha \phi^{\prime2}
\ee
We use the result (\ref{ssa}) to obtain 
\be
A=1-\alpha\,{\rm sech}^4(x)
\ee
and so, as before we change variable to write
\be
z=\int\frac{dx}{A}=x+\frac13\alpha\tanh(x)(2+{\rm sech}^2(x))
\ee
We invert this expression to get
\be
x=z-\frac13\alpha\tanh(z)(2+{\rm sech}^2(z))
\ee
The potential $U(z)$ which appears in the Schr\"odinger-like equation has the form
\ben
U(z)&=&4-6\,{\rm sech}^2(z)+\alpha\,{\rm sech}^2(z)\times\nonumber
\\
&&\,[15\,{\rm sech}^4(z)-11\,{\rm sech}^2(z)-2]
\een

In this case, since $\alpha$ is small we can write
\be
S^\dag S\, \eta=w^2\,\eta
\ee
where
\be
S=-\frac{d}{dz}+u(z)
\ee
and
\ben
u(z)&=&-2 \tanh(z)+\nonumber
\\
&&\frac13\alpha \tanh(z)\,{\rm sech}^2(z)(1+5\,{\rm sech}^2(z))
\een
which ensures stability of the defect solutions.

\section{Ending comments}
\label{3d}

The new defect solutions found in the former Sec.~\ref{scmodel} present features which may be of importance in applications. To elaborate on an interesting possibility, let us concentrate on the energy of domain walls generated by the defect solutions found in Sec.~\ref{scmodelII}. We first consider the standard situation, but now we work in higher spatial dimensions. To be specific, however, let us focus on the case of three spatial dimensions. Here the energy of static field configuration is
\be
E=\int\, d^3x\left(\frac12 (\nabla\phi)^2+V(\phi)\right)
\ee
We follow \cite{DH} to write $E_\lambda=K/\lambda+P/\lambda^3,$ where $\lambda$ is the scaling parameter and $K>0$ and $P>0$ stand for the kinetic and potential energies, respectively. The required minimization $dE_\lambda/d\lambda\to0$ for $\lambda\to1$ leads to $K+3P=0,$ and this shows that there is no stable defect solutions in this case. This result is old, but it guide us toward the new possibility which we now describe.   

We may circumvent the above reasoning with the extensions already investigated, which may contribute to stabilize the higher dimensional defect solutions. Although it is possible to make the investigation more general, for simplicity we will concentrate on the example described by (\ref{example}). In this case, the energy of the static field is given by
\be
E=\int\,d^3x\left(\frac12(\nabla\phi)^2+\alpha (\nabla\phi)^{4}+V(\phi)\right)
\ee 
and now the quantity $E_\lambda=K/\lambda+P/\lambda^3+{\bar K}\lambda$ may not collapse anymore, due to the new term ${\bar K}$ which represents the contribution added in the extended model. Indeed, the new defect structure may be stable, collapse or expand, depending on the contribution ${\bar K}$ being equal, lesser or greater than $K+3P.$ This new result shows very clearly that the extension considered in (\ref{example}) changes the standard scenario, leading to defect solutions which may engender distinct time evolution, of direct interest to applications in cosmology.

We illustrate the issue supposing that the defect solution presents spherical symmetry. In this case, we can think of it as a spherical domain wall with a given thickness, which can be identified with the width of the kinklike solution found in $(1,1)$ spacetime dimensions. As we have shown in Sec.{\ref{scmodelII}}, the term added in the extended model changes the width of the solution, and this may introduce additional effects into the time evolution of the defect structure in higher dimensions. A clear possibility concerns dependence of the energy density of the wall on $R,$ the radius of the spherical wall, and this may also change the way the defect evolves in time. A similar situation is described by an inflating elastic ball: as the ball inflates, the thickness of the elastic membrane decreases with increasing $R;$ if the total energy of the ball remains constant, its energy density $\sigma$ should depend on $R^{-2}$ to compensate the variation of the area of the ball.

In general, we write the energy density of the spherical domain wall as $\sigma=\sigma(R).$ We consider the relativistic case and we write the energy of the defect solution of radius $R$ in the form $E=\sigma(R)R^2/\sqrt{1-{\dot R}^2}.$ We suppose that $\sigma(R)\sim 1/R^a,$ $a$ real, to get 
\be
{\dot R}^2=1-\mu R^{2(2-a)}
\ee 
where $\mu$ is a parameter which depends on the initial velocity of the defect. We notice that $a=2$ leads to vanishing acceleration, and we get accelerated expansion for $a>2$ and collapse for $a<2.$ The case $a=0$ describes the situation in which the energy density of the wall does not depend on $R,$ which we recognize as the standard scenario -- see, for instance, Sec.~III of Ref.~\cite{sik} for more details on this issue. Thus, for $a\in(2,0)$ the defect collapses slower than in the standard scenario, but for $a\in(0,-\infty)$ it collapses as faster as lower is $a.$

Another line of investigation is related to the presence of two or more real scalar fields. The presence of more fields leads to two distinct classes of models, living in one or more spatial dimensions. In the case of a single spatial dimension, the presence of another field may contribute to add internal structure the the defect solution generated by the first field, as examined for instance in Refs.~{\cite{internal}}. In the case of two spatial
dimensions, we can use two fields to generate junctions of defects, as examined for instance in Refs.~\cite{junction1,junction2}. These extensions together with the extensions examined in the present work will certainly generate new models and new possibilities of internal structures and junctions.

To summarize, in this work we have investigated two distinct classes of models described by a single real scalar field in $(1,1)$ space-time dimensions.
In the first class of models, we have generalized the tachyonic dynamics, searching for the presence of static solutions, together with the corresponding stability profile. In the other case, for the second class of models we have extended the standard dynamics, investigating the presence of stable defect solutions in diverse contexts. We have constructed explicit defect structures, together with the specific considerations concerning stability. In particular, we have shown how some extensions may modify the standard evolution of domain walls in higher spatial dimensions, leading to distinct evolutions which are of direct interest to applications in diverse contexts.

We would like to thank CAPES, CNPq and PRONEX/CNPq/FAPESQ for financial support. JCREO is specially grateful to Department of Physics, Federal University of Para\'\i ba, for the kind hospitality during the beginning of the investigations.

\end{document}